\documentclass[12pt,preprint]{aastex}
\usepackage{natbib}
\usepackage{amsmath}
\usepackage{rotating}
\usepackage{gensymb}
\usepackage{lscape}

\newcommand{\Msun}{$M_{\odot}$}
\newcommand{\dmu}{pc cm$^{-3}$}
\newcommand{\us}{$\mu$s}

\newcommand{\Pdot}{\textit{\.{P}}}

\shorttitle{The Pulsar Search Collaboratory: Discovery and Timing of Five New Pulsars}
\shortauthors{Rosen, Heatherly, Lorimer, Lynch, McLaughlin, McLaughlin, Kondratiev, Ransom, Yun}

\begin{document}

\title{The Pulsar Search Collaboratory: Discovery and Timing of Five New Pulsars}

\author{R. Rosen}
\affil{West Virginia University, White Hall, Morgantown, WV 26506}
\email{Rachel.Rosen@mail.wvu.edu }
\author{J. Swiggum}
\affil{West Virginia University, White Hall, Morgantown, WV 26506}
\author{M. A. McLaughlin}
\affil{West Virginia University, White Hall, Morgantown, WV 26506}
\affil{Adjunct Astronomer, National Radio Astronomy Observatory}
\author{D. R. Lorimer}
\affil{West Virginia University, White Hall, Morgantown, WV 26506}
\affil{Adjunct Astronomer, National Radio Astronomy Observatory}
\author{M. Yun}
\affil{West Virginia University, White Hall, Morgantown, WV 26506}
\author{S. A. Heatherly}
\affil{NRAO, P.O. Box 2, Green Bank, WV 24944}
\author{J. Boyles}
\affil{West Virginia University, White Hall, Morgantown, WV 26506}
\affil{Western Kentucky University, Bowling Green, KY 42101}
\author{R. Lynch}
\affil{McGill University, Rutherford Physics Building, 3600 Rue University, Montreal, QC H3A 2T8}
\author{V. I. Kondratiev}
\affil{ASTRON, the Netherlands Institute for Radio Astronomy, Postbus 2, 7990 AA, Dwingeloo, The Netherlands}
\affil{Astro Space Center of the Lebedev Physical Institute, Profsoyuznaya str. 84/32, Moscow 117997, Russia}
\author{S. Scoles}
\affil{NRAO, P.O. Box 2, Green Bank, WV 24944}
\author{S. M. Ransom}
\affil{NRAO, 520 Edgemont Road, Charlottesville, VA 22903}
\author{M. L. Moniot}
\affil{James River High School, 9906 Springwood Road, Buchanan, VA  24066}
\author{A. Cottrill}
\affil{Lincoln High School, 100 Jerry Toth Drive, Shinnston, WV 26431}
\author{M. Weaver}
\affil{Broadway High School, 269 Gobbler Drive, Broadway, VA 22815}
\author{A. Snider}
\affil{Sherando High School, 185 South Warrior Drive, Stephens City, VA 22655}
\author{C. Thompson}
\affil{James River High School, 9906 Springwood Road, Buchanan, VA  24066}
\author{M. Raycraft}
\affil{Lincoln High School, 100 Jerry Toth Drive, Shinnston, WV 26431}
\author{J. Dudenhoefer}
\affil{Hedgesville High School, 109 Ridge Road North, Hedgesville, WV 25427}
\author{L. Allphin}
\affil{Hedgesville High School, 109 Ridge Road North, Hedgesville, WV 25427}
\author{J. Thorley}
\affil{Strasburg High School, 250 Ram Drive, Strasburg, VA 22657} 
\author{B. Meadows}
\affil{Independence High School, 850 Independence Road, P.O. Box 1595, Coal City, WV 25823}
\author{G. Marchiny}
\affil{Lincoln High School, 100 Jerry Toth Drive, Shinnston, WV 26431}
\author{A. Liska}
\affil{Hedgesville High School, 109 Ridge Road North, Hedgesville, WV 25427}
\author{A. M. O'Dwyer}
\affil{Nicolet High School, 6701 N. Jean Nicolet Road, Glendale, WI 53217}
\author{B. Butler}
\affil{Lincoln High School, 100 Jerry Toth Drive, Shinnston, WV 26431}
\author{S. Bloxton}
\affil{Nicholas County High School, 30 Grizzly Road, Summersville, WV 26651}
\author{H. Mabry}
\affil{Rowan County High School, 499 Viking Drive, Morehead, KY  40351}
\author{H. Abate}
\affil{James W. Robinson Secondary School, 5035 Sideburn Road Fairfax, VA 22032}
\author{J. Boothe}
\affil{Pocahontas County High School, Route 1 Box 133A, Dunmore, WV 24934}
\author{S. Pritt}
\affil{Elkins High School, 100 Kennedy Drive, Elkins, WV  26241}
\author{J. Alberth}
\affil{Central High School of Westosha, 75th Street, P.O. Box 38, Salem, WI 53168}
\author{A. Green}
\affil{Hedgesville High School, 109 Ridge Road North, Hedgesville, WV 25427}
\author{R. J. Crowley}
\affil{Trinity High School, 231 Park Avenue, Washington, PA 15301} 
\author{A. Agee}
\affil{Roanoke Valley Governor's School, 2104 Grandin Rd SW, Roanoke, VA 24015}
\author{S. Nagley}
\affil{Hedgesville High School, 109 Ridge Road North, Hedgesville, WV 25427}
\author{N. Sargent}
\affil{Independence High School, 850 Independence Road, P.O. Box 1595, Coal City, WV 25823}
\author{E. Hinson}
\affil{Blacksburg High School, 3109 Price's Fork Road, Blacksburg, VA 24060}
\author{K. Smith}
\affil{Nicholas County High School, 30 Grizzly Road, Summersville, WV 26651}
\author{R. McNeely}
\affil{Blacksburg High School, 3109 Price's Fork Road, Blacksburg, VA 24060}
\author{H. Quigley}
\affil{Wilmington Middle School, 715 S. Joliet Street, Wilmington, IL 60481}
\author{A. Pennington}
\affil{Tolsia High School, 1 Rebel Drive, Fort Gay, WV 25514}
\author{S. Chen}
\affil{Blacksburg High School, 3109 Price's Fork Road, Blacksburg, VA 24060}
\author{T. Maynard}
\affil{Tolsia High School, 1 Rebel Drive, Fort Gay, WV 25514}
\author{L. Loope}
\affil{James River High School, 9906 Springwood Road, Buchanan, VA  24066}
\author{N. Bielski}
\affil{Central High School of Westosha, 75th Street, P.O. Box 38, Salem, WI 53168}
\author{J. R. McGough}
\affil{Strasburg High School, 250 Ram Drive, Strasburg, VA 22657} 
\author{J. C. Gural}
\affil{Trinity High School, 231 Park Avenue, Washington, PA 15301} 
\author{S. Colvin}
\affil{Spring Valley High School, 1 Timberwolf Drive, Huntington, WV 25704}
\author{S. Tso}
\affil{James W. Robinson Secondary School, 5035 Sideburn Road Fairfax, VA 22032}
\author{Z. Ewen}
\affil{James River High School, 9906 Springwood Road, Buchanan, VA  24066}
\author{M. Zhang}
\affil{Morgantown High School, 109 Wilson Avenue, Morgantown, WV 26501}
\author{N. Ciccarella}
\affil{James River High School, 9906 Springwood Road, Buchanan, VA  24066}
\author{B. Bukowski}
\affil{Sherando High School, 185 South Warrior Drive, Stephens City, VA 22655}
\author{C. B. Novotny}
\affil{Hedgesville High School, 109 Ridge Road North, Hedgesville, WV 25427}
\author{J. Gore}
\affil{Morgantown High School, 109 Wilson Avenue, Morgantown, WV 26501}
\author{K. Sarver}
\affil{Spring Valley High School, 1 Timberwolf Drive, Huntington, WV 25704}
\author{S. Johnson}
\affil{Rowan County High School, 499 Viking Drive, Morehead, KY  40351}
\author{H. Cunningham}
\affil{Roanoke Valley Governor's School, 2104 Grandin Rd SW, Roanoke, VA 24015}
\author{D. Collins}
\affil{Lincoln High School, 100 Jerry Toth Drive, Shinnston, WV 26431}
\author{D. Gardner}
\affil{Blacksburg High School, 3109 Price's Fork Road, Blacksburg, VA 24060}
\author{A. Monteleone}
\affil{Morgantown High School, 109 Wilson Avenue, Morgantown, WV 26501}
\author{J. Hall}
\affil{Rowan County High School, 499 Viking Drive, Morehead, KY  40351}
\author{R. Schweinhagen}
\affil{Carmel Catholic High School, One Carmel Parkway, Mundelein, IL 60060}
\author{J. Ayers}
\affil{James River High School, 9906 Springwood Road, Buchanan, VA  24066}
\author{S. Jay}
\affil{Wilmington Middle School, 715 S. Joliet Street, Wilmington, IL 60481}
\author{B. Uosseph}
\affil{James W. Robinson Secondary School, 5035 Sideburn Road Fairfax, VA 22032}
\author{D. Dunkum}
\affil{Spring Valley High School, 1 Timberwolf Drive, Huntington, WV 25704}
\author{J. Pal}
\affil{Rowan County High School, 499 Viking Drive, Morehead, KY  40351}
\author{S. Dydiw} 
\affil{Trinity High School, 231 Park Avenue, Washington, PA 15301} 
\author{M. Sterling} 
\affil{Langley High School, 6520 Georgetown Pike, McLean, VA 22101}
\author{E. Phan} 
\affil{George C. Marshall High School, 7731 Leesburg Pike, Falls Church, VA 22043} 
\begin{abstract} 

We present the discovery and timing solutions of five new pulsars by students involved in the Pulsar Search Collaboratory (PSC), a NSF-funded joint program between the National Radio Astronomy Observatory and West Virginia University designed to excite and engage high-school students in Science, Technology, Engineering, and Mathematics (STEM) and related fields. We encourage students to pursue STEM fields by apprenticing them within a professional scientific community doing cutting edge research, specifically by teaching them to search for pulsars. The students are analyzing 300 hours of drift-scan survey data taken with the Green Bank Telescope at 350 MHz. These data cover 2876 square degrees of the sky. Over the course of five years, more than 700 students have inspected diagnostic plots through a web-based graphical interface designed for this project. The five pulsars discovered in the data have spin periods ranging from 3.1 ms to 4.8 s. Among the new discoveries are \---  PSR J1926$-$1314, a long period, nulling pulsar; PSR J1821+0155, an isolated, partially recycled 33-ms pulsar; and PSR J1400$-$1438, a millisecond pulsar in a 9.5-day orbit whose companion is likely a white dwarf star.

\end{abstract}

\section{Introduction}

Pulsars, rapidly rotating highly magnetized neutron stars, are most commonly studied at radio wavelengths. The first pulsar was discovered in 1967 \citep{hew68} and currently over 2000 of these objects have been cataloged\footnote{For an up-to-date catalog of pulsars, see \url {http://www.atnf.csiro.au/research/pulsar/psrcat}} \citep{man05}. Over 30,000 normal pulsars are potentially detectable in the galaxy (i.e. those beaming towards us) \citep{lor06} and pulsar surveys covering larger sky area with the sensitivity achieved with modern hardware and telescopes are the best means of detecting a larger sample of the pulsar population.

Continued pulsar searches are crucial for finding exotic objects, which include pulsars with ultra-fast spins (known as millisecond pulsars or MSPs; \citet{bac82}) or in very relativistic orbits, binary pulsar systems, and young high velocity pulsars. Binary systems are interesting laboratories for new physics, whether the companion is another neutron star, providing tests of general relativity \citep{lyn04a,kra04,bur03,hul75}, a white dwarf star, giving insight into white dwarf cooling \citep{dri98}, or a main sequence star, probing stellar winds \citep{mad12}.  The discovery of young, high velocity pulsars can shed light on the supernova explosions in which pulsars are formed \citep{arz02}. In addition, some pulsars exhibit intermittent emission, such as Rotating Radio Transients (RRATs) \citep{mcl06} or extreme nulling \citep{li12,esa05} which give insight into the pulsar magnetosphere and how the magnetosphere affects the spin-down rate of the pulsar \citep{lyn10,kra06b}.

Pulsar searches are also necessary for finding MSPs which can be used for gravitational wave detection. MSPs are old pulsars that are thought to have gained their rapid spins during a process of accretion from a companion star \citep{alp82}. Because of their extremely high rotational precision, MSPs may be able to be used to directly detect gravitational waves. A stochastic gravitational wave background, single, continuous gravitational wave sources, burst sources, or burst sources with memory can be detected through pulsar timing arrays \citep[PTAs;][]{fos90}. The success of PTAs depends crucially on the number of MSPs in the array and the sensitivity also increases with more MSPs distributed throughout the sky to sample the whole range of angles between cross-correlation pairs \citep{jen06}. 

The 100-m Robert C. Byrd Green Bank Telescope (GBT)\footnote{The Robert C. Byrd Green Bank Telescope (GBT) is operated by the National Radio Astronomy Observatory which is a facility of the U.S. National Science Foundation operated under cooperative agreement by Associated Universities, Inc.} is one of the best telescopes for pulsar searching because of its large collecting area, sky coverage ability, state-of-the-art data acquisition hardware, and location in a radio quiet zone. The GBT 350-MHz Drift Scan Survey, completed during the summer of 2007 while the GBT was under track replacement, comprised over 1491 hours of data totaling 134 TB \citep[hereafter Papers I and II, respectively;][]{boy12,lyn12}. Of these data, the Pulsar Search Collaboratory (PSC) received 300 hours, approximately 30 TB of data. The Pulsar Search Collaboratory is a partnership between the National Radio Astronomy Observatory (NRAO) and West Virginia University (WVU) designed to engage students in Science, Technology, Engineering, and Mathematics (STEM) and related fields by using information technology to conduct current scientific research, specifically searching for new pulsars.

One of the limiting factors in discovering radio pulsars is the large volume of data and the ability to analyze the diagnostic plots quickly. The data are processed through a software pipeline on computing clusters, searching for both periodic and transient signals. The resulting diagnostic plots are visually inspected, in this case, by students of the PSC. In its current form, plot inspection requires a basic understanding of pulsars and many person-hours. While the human eye is remarkably adept at distinguishing a signal from noise, the next generation of pulsar data analysis may require intelligent algorithms, such as neural nets \citep{eat10}, due to the large amounts of data produced.
 
In this paper, we outline the data collection and processing of the entire GBT 350-MHz Drift Scan Survey and how the data for the PSC relate to the data acquired for the astronomical community in \S\ref{data}. We discuss the PSC program in \S\ref{psc}. In \S\ref{results}, we highlight the scientific discoveries made by the students and discuss these results in \S\ref{discussion}.

\section{Data Collection and Processing}
\label{data}

The data for the GBT 350-MHz Drift Scan Survey were acquired during May though August of 2007 while the GBT underwent a track replacement. The telescope was parked at an azimuth of $\sim229^\circ$ for the first half of the observations and at $\sim192^\circ$ for the remainder of the observations and covered declinations of $-8^\circ \lesssim \delta \lesssim +38^\circ$ and $-21^\circ \lesssim \delta \lesssim +38^\circ$, depending on azimuth (Paper I).  The data were collected as the sky drifted overhead and divided into pseudo-pointings, where a pseudo-pointing is a continuous chunk of data $\sim140$ seconds in length, equal to the transit time through the center of the beam of a source at 350 MHz. Each pseudo-pointing overlaps the preceding one by 50\%, so each piece of data is processed twice. This retains sensitivity to sources which may have fallen on the edge of a pointing (Paper II).

\begin{figure}
\begin{center}
\includegraphics[scale=0.6,angle=-90]{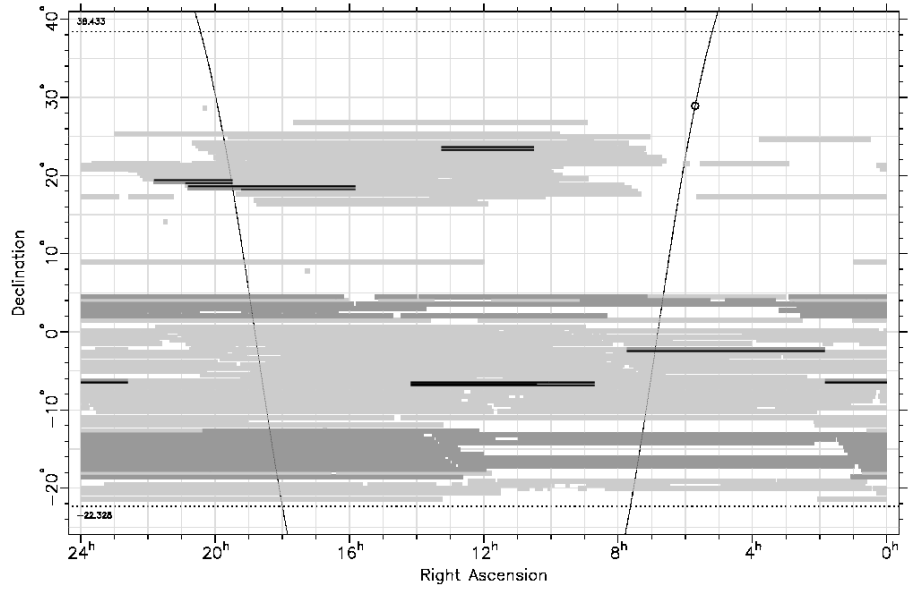}
\end{center}
\caption{The portions of the sky covered by the PSC (grey), the astronomy community (light grey), and where the astronomy community and PSC overlap (black).  The line represents the Galactic plane and the circle is the location of the Galactic anticenter. Some portions of the sky were covered twice and the overlap of the PSC data with the astronomy community is due to the distribution of data.}
\label{fig:pscsky_overlap}
\end{figure}

The PSC received approximately 300 hours of the GBT 350-MHz Drift Scan Survey data. Each hour of observations results in about 48 pseudo-pointings separated by about 30 arcminutes in right ascension, resulting in approximately 16,500 pseudo-pointings.  In total, the PSC data covered 2,876 square degrees of the sky.  Figure \ref{fig:pscsky_overlap} shows the overlap between the portions of the sky searched by the PSC compared to the rest of the GBT 350-MHz Drift Scan Survey.
  
The survey was conducted at 350 MHz because of the low radio frequency interference at the telescope at this frequency, the steep spectral indices of most pulsars \citep{sie73}, and the large ($\sim36'$) beam size of the GBT which allows for a greater amount of sky coverage than higher frequency receivers (Paper I). The Spigot autocorrelation spectrometer \citep{kap05} was used to acquire the data during the observations with 50 MHz of bandwidth at 81.92 $\mu$s sampling time with 2048 channels and 8 bit precision.

Each pseudo-pointing was processed using the {\sc PRESTO}\footnote{\url {http://www.cv.nrao.edu/}$\sim$\url{sransom/presto/}} software package \citep{ran01}. {\sc PRESTO} performs a series of Fourier transforms to search for periodicity, including harmonic summing to improve sensitivity to narrow pulses and allowing constant accelerations for pulsars in binary orbits \citep{lor05}.  In addition, single-pulse searches for bright individual pulses are conducted. The dispersion measure (DM) is the integrated column density of electrons along the line of sight to the pulsar. Because the DM is unknown, both types of searches are made for 9700 trial values of the DM up to 1015.5 \dmu.  The periodicity searches are most sensitive to normal, periodic  pulses, while the single-pulse search is designed to detect bursty, sporadic sources like rotating radio transients \citep{mcl06}.  {\sc PRESTO} creates the periodicity search diagnostic plots by folding the raw data modulo the pulse period and then optimizing over local values for dispersion, period, and period-derivative. The students inspect the top 30 candidates output from {\sc PRESTO} for each beam and we use {\sc PRESTO} to create single-pulse search plots in five different DM ranges. The details of the processing pipeline can be found in Paper II.

After the data have been processed, the students analyze the diagnostic pulsar plots using a quantitative grading scheme.  For each pseudo-pointing, 35 diagnostic plots (30 periodicity search plots and five single-pulse search plots) are placed in the PSC database\footnote{\url {http://psrsearch.wvu.edu}}. The database contains approximately 1000$-$1500 pointings visible at any given time.  Only one student from each school can score a given pseudo-pointing \citep{ros_aer}.  Each pseudo-pointing must be scored by students on five individual teams for redundancy, after which the pseudo-pointing is no longer visible in the database and it is submitted to the astronomers for consideration.  This system is designed to both prevent weak pulsars from escaping detection while also keeping the number of potential new candidates to a manageable number.  Only those pointings with top rankings by students are reviewed by the astronomers and considered for follow-up observations.

\section{PSC Program}
\label{psc}

A full description of the PSC program and its educational impact on students and STEM careers is detailed in \citet{ros_aer}. The PSC is currently in its fifth year and we now have 76 teachers and 705 students from 18 states participating. From both a scientific and educational perspective, the PSC has been a success. In addition to the scientific discoveries outlined in this paper, the PSC program has reached significant educational goals. These are discussed in \citet{ros_aer} and the highlights are: the PSC is reaching low-income students and attracts students who are first generation college-goers; the PSC succeeds in building confidence in students, rapport with the scientists involved in the project, and greater comfort with teamwork; we see additional gains in girls as they see themselves more as scientists after participating in the PSC program. This result is exciting and significant as self-efficacy is an important predictor of success in STEM fields.

Of the approximately 16,500 pseudo-pointings in the PSC data, the students have completely analyzed (specifically, the pseudo-pointing has been examined by five different students on five different teams) 6,600 pseudo-pointings, or about 40\% of the data. In the analyzed data, the students have discovered five new pulsars, on average one new pulsar per 1,320 pseudo-pointings, where a given pulsar can be discovered by multiple students. By comparison, the astronomical community analyzed approximately 61,200 pseudo-pointings in their drift-scan data, discovering 31 new pulsars (see Papers I and II), resulting in a detection rate of one new pulsar per 1,900 pseudo-pointings. Given the random distribution of data between the astronomers and the PSC, and the random pseudo-pointings through which the students have already analyzed, the detection rate of pulsars from the PSC students is on par with the astronomers.

\section{Results}
\label{results}

\subsection{Known Pulsars}

The presence of known pulsars in the PSC data allows us to test the survey sensitivity and to quantify the students' ability to detect pulsars by comparing the number of known pulsars expected to be detected, based on survey area and sensitivity to the known pulsars flagged by the students. To determine the number of known pulsars detectable in the PSC data, we calculate the survey sensitivity for a given frequency ($\nu$) as a function of the phase-averaged limiting flux density \citep[Paper II]{dew85,lor05}:

\begin{equation}
S_{\nu,\rm min} = \frac{\beta}{\epsilon}\frac{({\rm S/N}_{\rm min})T_{\rm sys}}{G\sqrt{{n_{\rm p}}{t_{\rm int}}\Delta\nu}}\sqrt{\frac{W}{P-W}},
\label{eqn:sensitivity}
\end{equation}
\\where $\beta = 1.2$ is a correction factor accounting for digitization losses \citep{lor05}, $\epsilon$ accounts for the drift of the source and any degradation of the signal if the source does not drift precisely through the beam center (Paper II), S/N$_{\rm min}$ is the signal-to-noise threshold, $T_{\rm sys}$ (K) is the combined system and sky temperature, $G$ (K Jy$^{-1}$) is the telescope gain, $n_{\rm p}$ is the number of summed polarizations, $t_{\rm int}$ (s) is the observation length, $\Delta\nu$ (MHz) is the bandwidth, and $W$ is the pulse width. 

\begin{figure}
\begin{center}
\includegraphics[scale=1]{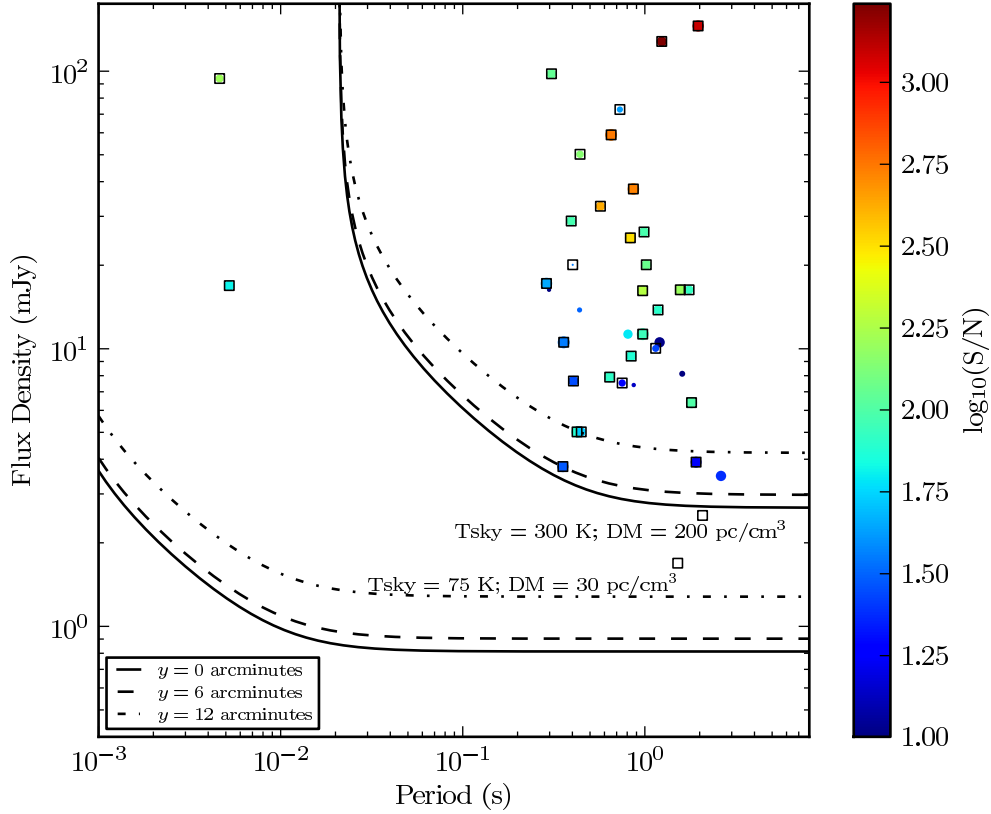}
\end{center}
\caption{The students' detection of known pulsars. The known pulsars in the area of the sky covered by the PSC data are shown as colored points as a function of flux density where the detection threshold is S/N $\geq$~9. The 350-MHz fluxes are calculated by assuming a spectral index of $-1.7$ \citep{mar00} and extrapolating from 400 MHz (or 1400 MHz if no 400 MHz values are listed) values taken from the ATNF catalog. Note that some scatter is expected as the observed flux of the pulsars in the PSC data may vary due to scintillation. The size of the point corresponds to the closest approach angle to the center of the beam with larger points being closer to the center. The dashed lines are sensitivity curves based on Equation \ref{eqn:sensitivity} for various sky temperatures and DMs at different distances $y$ from the beam center. Squares represent the known pulsars that the PSC students have detected.}
\label{fig:sensitivity}
\end{figure}

Figure \ref{fig:sensitivity} shows the students' detection of known pulsars. Each colored point represents a known pulsar within the sky area covered by the PSC data.  The size of the point corresponds to the closest approach angle to the center of the 350-MHz beam with larger points being closer to the center. The signal-to-noise (S/N) is shown in color where we expect the students to detect known pulsars with a S/N $\geq$~9. The dashed lines are sensitivity curves which incorporate sky temperature and DM for different distances $y$ from the beam center.

The black squares are known pulsars that the PSC students have detected. In the data that the students have analyzed, there are 39 known pulsars predicted to be detectable above a S/N threshold of nine. Of these, the students found 32 of them as shown by the colored points outlined with a black square. The seven pulsars that were undetected by students (colored points without black squares) were likely not detectable due to strong radio frequency interference, scintillation, scattering, or diminished sensitivity at the edges of the telescope beam. These pulsars are also not detectable when we fold the raw data with a known ephemeris. In addition, the students have detected two known pulsars below the S/N $=$~9 threshold, represented by the two squares without colored points, and three other pulsars which did not have any recorded flux measurements and are therefore not plotted in Figure \ref{fig:sensitivity}. We find that the students correctly found and identified all the known, detectable pulsars in the data they have examined.

\subsection{New Pulsars}
\label{newpulsars}

Once a potential pulsar was detected by the students, we conducted a confirmation observation at 350 MHz with the GBT. If detected at 350 MHz, we then followed with a second observation at 820 MHz. Because of the smaller beam size at 820 MHz, we took a series of gridding observations to determine a more precise location \citep{mor02}. Gridding entails observing a set of locations at 820 MHz which covers the area of the original 350 MHz beam to find the Right Ascension and Declination where the pulsar has its maximum intensity. Upon confirmation at 820 MHz, timing observations commenced for each new PSC pulsar. Timing solutions provide a more precise location of the pulsar and allow us to measure and/or characterize a variety of pulsar properties, such as period, period derivative, proper motion, and any orbital parameters, should the pulsar be in a binary system (Paper II). All gridding observations were conducted at 820 MHz using the Green Bank Ultimate Pulsar Processing Instrument \citep[GUPPI;][]{dup08} with 200 MHz bandwidth at 81.92 $\mu$s time sampling resolution. Timing observations for all the pulsars except the MSP were conducted at 820 MHz in the same configuration as the gridding observations. For the MSP, we included some timing observations at 350 MHz using GUPPI with 100 MHz bandwidth at 81.92 $\mu$s time sampling resolution.

Most of the PSC pulsar timing campaigns began with a series of closely spaced observations followed by monthly observations. Each observation lasts approximately 10--20 minutes. From these observations, standard pulse profiles were created by summing data over all the frequency channels within the 200 MHz bandwidth. If the observation was at 350 MHz, as in the case of the MSP, we used a different template. The standard profiles were then used to create pulse times-of-arrival (TOAs) using {\sc PSRCHIVE}\footnote{\url {http://psrchive.sourceforge.net}}\citep{hot04}. We follow the procedure for the timing analysis outlined in Paper I, where phase connected timing solutions were created using {\sc TEMPO2}\footnote{\url {http://tempo2.sourceforge.net}} and the DE405 Solar System ephemeris\footnote{\url {http://iau-comm4.jpl.nasa.gov/de405iom/}}, with the modification that we create three TOAs per epoch for all pulsars if possible, not just binary systems. All of our timing solutions are referenced to UTC(NIST). Global EFACs (multiplicative factors) have been applied to individual TOA error bars such that the resulting $\chi^2$ value is one. The parameter uncertainties shown in Tables \ref{tab:slow_pulsars} and \ref{tab:msp} are double the quantities reported by {\sc TEMPO2}, giving a conservative estimate of 68\% uncertainties.

To calculate the flux, we took calibration scans before and after each observation at each frequency.  Using the flux calibrator 3C190 measured at both 350 MHz and 820 MHz, we created flux calibrated profiles using {\sc PSRCHIVE}. For those observations with off-pulse flux due to radio frequency interference, we manually estimated the on and off pulse regions and used these to calculate the mean flux.

\begin{figure}
\begin{center}
\includegraphics[scale=1]{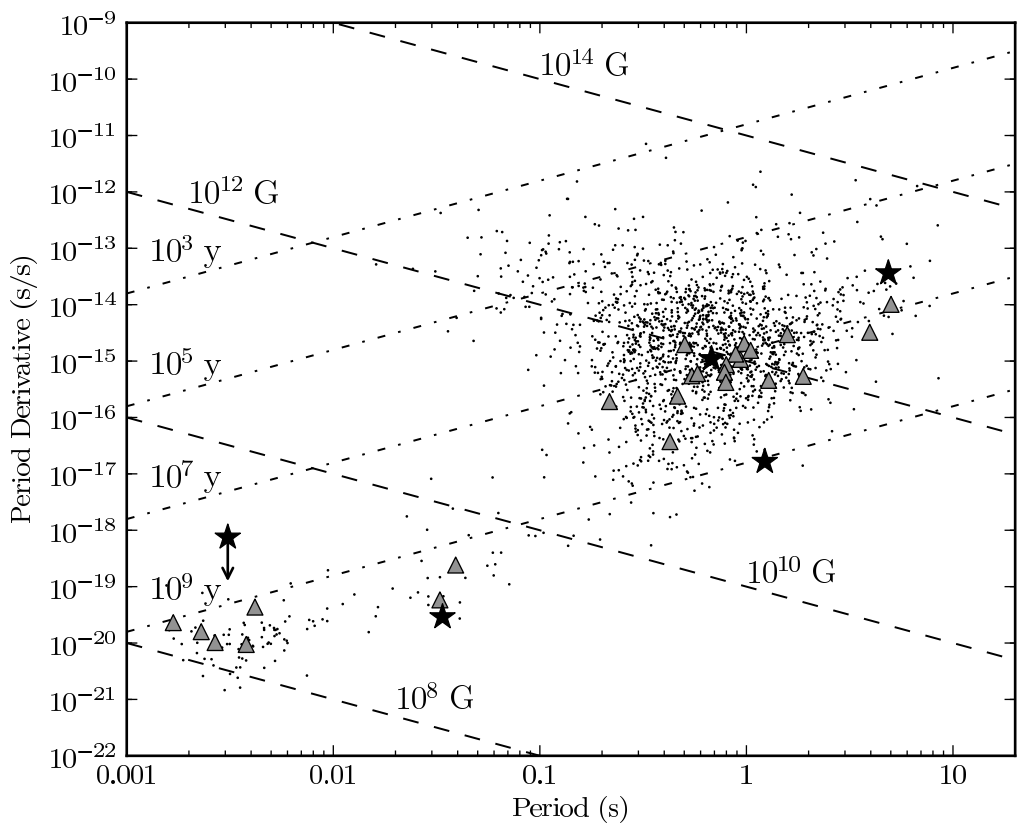}
\end{center}
\caption{Periods and period derivatives of all known pulsars not in globular clusters(dots), those discovered in the GBT 350-MHz Drift Scan Survey (triangles), and those discovered by the PSC (stars). The star representing the MSP, PSR J1400$-$1438, has been placed at the current upper limit for its period derivative, since we have not yet been able to measure this quantity precisely. The dashed lines represent constant magnetic field strength and constant characteristic age \citep[e.g.][]{lor05}.}
\label{fig:ppdot}
\end{figure}

\begin{figure}
\begin{center}
\includegraphics[scale=1]{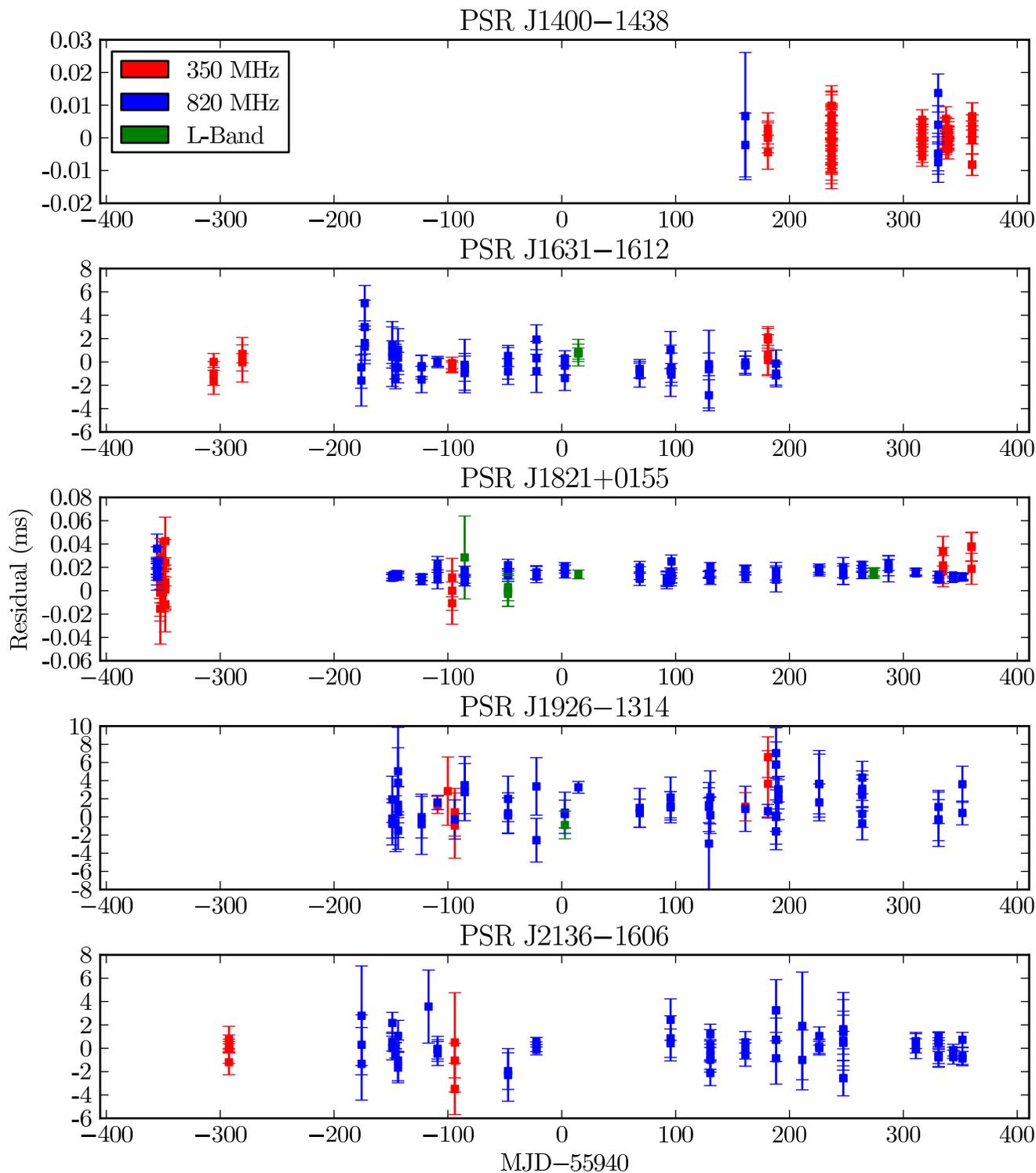}
\end{center}
\caption{Timing residuals for each of the newly discovered pulsars by the PSC students. All TOAs are shown for the slow pulsars and they are all phase-connected. For PSR J1400$-$1438, we have omitted the original 820 MHz TOAs.}
\label{fig:residuals}
\end{figure}

In the data that they have examined so far, the students have discovered five new pulsars, one of which is an MSP. We are conducting timing observations on all of them and they are shown on the $P$-\Pdot~ diagram in Figure \ref{fig:ppdot}. The pulsars discovered by the PSC are indicated with stars and those discovered in the GBT 350-MHz Drift Scan Survey are shown with triangles. The timing parameters of the four normal pulsars, including the root mean square (RMS) residual -- the difference between when we expect a pulse to arrive and when it actually does -- are outlined in Table \ref{tab:slow_pulsars}. Because the MSP, PSR J1400$-$1438, is in a binary system, we discuss its timing and orbital parameters in \S\ref{msp}. The pulse profiles and timing residuals of the pulsars are shown in Figures \ref{fig:slow_profs} and \ref{fig:residuals}, respectively.

\begin{landscape}
\begin{deluxetable}{lcccc}
  \centering
  \tabletypesize{\footnotesize}
  \tablewidth{0pt}
  \tablecolumns{5}
  \tablecaption{Parameters of Newly Discovered Pulsars}
  \tablehead{
    \colhead{Parameter} &
    \colhead{PSR J1631$-$1612} &
    \colhead{PSR J1821+0155} &
    \colhead{PSR J1926$-$1314} &
    \colhead{PSR J2136$-$1606}}
  \startdata
    \cutinhead{Spin \& Astrometric Parameters}
    Right Ascension (J2000) \dotfill & 16:31:52.47(3) & 18:21:38.88368(5) & 19:26:53.835(3) & 21:36:00.22(10) \\
    Declination (J2000) \dotfill & $-$16:12:52.0(3.0) & +01:55:21.993(2) & $-$13:14:03.8(1.8) & $-$16:06:13.0(4.0) \\
    Pulsar Period (s) \dotfill & 0.67768391249(4) & 0.03378133193189(13) & 4.86428379983(10) & 1.22723540015(6) \\
    Period Derivative (s s$^{-1}$) \dotfill & 1.096(4)$\times$10$^{-15}$ & 2.900(19)$\times$10$^{-20}$ & 3.6440(5)$\times$10$^{-14}$ & 1.6(3)$\times$10$^{-17}$ \\
    Dispersion Measure (\dmu) \dotfill & 33.77(2) & 51.75415(3) & 40.83(5) & 18.48(5) \\
    Reference Epoch (MJD) \dotfill & 55766.0 & 55584.0 & 55791.0 & 55764.0 \\
    Span of Timing Data (MJD) \dotfill & 55634--56128 & 55584--56299 & 55150--56291 & 55647--56291 \\
    Number of TOAs \dotfill & 65 & 122 & 75 & 71 \\
    RMS Residual (\us) \dotfill & 820.636 & 6.396 & 1583.625 & 652.733 \\
    EFAC \dotfill & 2.8724 & 1.3291 & 1.6556 & 1.9277 \\
    \cutinhead{Derived Parameters}
    Surface Magnetic Field ($10^{12}$ Gauss) \dotfill & 0.8721 & 0.001002 & 13.5 & 0.1441 \\
    Spin-down Luminosity ($10^{32}$ erg s$^{-1}$) \dotfill & 1.39 & 0.297 & 0.125 & 0.00353 \\
    Characteristic Age (Myr) \dotfill & 9.8 & 18500.0 & 2.12 & 1180.0 \\
    Distance (kpc) \dotfill & 1.3 & 1.9 & 1.5 & 0.8 \\
 \enddata
\label{tab:slow_pulsars}
\end{deluxetable}
\end{landscape}

\subsubsection{PSR J1400$-$1438}
\label{msp}

PSR J1400$-$1438 is the only MSP discovered in the PSC portion of the drift-scan data to date. At a distance of 0.5 kpc \citep[DM of 4.6 \dmu;][]{cor02} with a flux of 4.6 mJy at 350 MHz, it is a bright, nearby pulsar. Only four other MSPs have lower DMs than PSR J1400$-$1438. At 350 MHz, we measure TOAs with accuracies of $\sim$1 $\mu$s  in one-minute integrations. According to several critera, such as pulse width and location, PSR J1400$-$1438 looks like a valuable addition to PTAs, but weak detections at 820 MHz (and no detections at higher frequencies) suggest that the object has a steep spectral index or scintillates havily, perhaps excluding it from consideration. Therefore, we have excluded TOAs from early followup observations at 820 MHz, often with error bars five times larger than TOAs from 350 MHz or coherent dedispersion observations at 820 MHz. More observations are required to determine whether this pulsar will be useful for gravitational wave detection and these will be conducted with coherent dedispersion to minimize profile smearing.

PSR J1400$-$1438 has an orbital period of 9.5 days and a projected semi-major axis of 8.4 lt-s, indicating that its companion is likely a white dwarf star with a minimum mass of 0.26 \Msun~(see Table \ref{tab:msp} for the orbital parameters). The minimum companion mass and orbital period for PSR J1400$-$1438 match with the published core-mass--orbital-period relation \citep{tau99}. \citet{tau99} show that the companion mass varies slightly for an orbital period, depending on the composition of the donor star. The minimum companion mass of PSR J1400$-$1438 is 0.26\Msun, which falls between projections for a Population I donor star (0.243\Msun) and a Population II donor star (0.267 \Msun).

The timing measurements are not complete, and thus we do not yet have an accurate spin-down rate. However, by fitting for the position and orbital parameters, we are able to calculate an upper limit on the spin-down rate, as shown in Figure \ref{fig:ppdot}.

\begin{deluxetable}{lc}
  \centering
  \tabletypesize{\footnotesize}
  \tablewidth{0pt}
  \tablecolumns{2}
 \tablecaption{Parameters of Newly Discovered MSP}
   \tablehead{
    \colhead{Parameter} &
     \colhead{PSR J1400$-$1438}}
  \startdata
  \cutinhead{Spin \& Astrometric Parameters}
    Right Ascension (J2000)                       \dotfill & 14:00:30.34(7)              \\
    Declination (J2000)                           \dotfill & $-$14:38:14.4(2.0)            \\
    Pulsar Period (s)                             \dotfill & 0.0030842332007(2)          \\
    Period Derivative (s s$^{-1}$)                       \dotfill & $<7.5\times10^{-19}$   \\
    Dispersion Measure (\dmu)                     \dotfill & 4.928(3)                    \\
    Reference Epoch (MJD)                         \dotfill & 56008.0                       \\
    Span of Timing Data (MJD)                     \dotfill & 56101--56300                  \\
    Number of TOAs                                \dotfill & 82                            \\
    RMS Residual (\us)                            \dotfill & 3.347                         \\
    EFAC                                          \dotfill & 1.1372                        \\
    \cutinhead{Binary Parameters}
    Orbital Period (days)                         \dotfill & 9.54743(2)                  \\
    Projected Semi-major Axis (lt-s)              \dotfill & 8.4211(3)                   \\
    Epoch of Periastron (MJD)                     \dotfill & 55955.90101(6)                \\
    \cutinhead{Derived Parameters}
    Orbital Eccentricity                          \dotfill & 0.000074(8)                   \\
    Longitude of Periastron (deg)                 \dotfill & 59.3(9.0)                     \\
    Epoch of Periastron (MJD)                     \dotfill & 55957.5(2)                  \\
    Surface Magnetic Field ($10^{12}$ Gauss)      \dotfill & $<$0.001541                   \\
    Spin-down Luminosity ($10^{32}$ erg s$^{-1}$)        \dotfill & $<$10100.0                    \\
    Characteristic Age (Myr)                      \dotfill & $>$65.0                       \\
    Distance (kpc) 				  \dotfill & 0.5				\\
    Mass Function (\Msun)                         \dotfill & 0.0070(6)                     \\
    Minimum Companion Mass (\Msun)                \dotfill & 0.2636                        \\
     \enddata
  \tablecomments{All timing solutions use the DE405 Solar System Ephemeris and the UTC(NIST) time system. We use the ELL1 timing model as it is preferable for low eccentricity orbits. The minimum companion mass is calculated assuming a pulsar mass of 1.35\Msun. Because we have less than one year of timing data, the position and the period derivative are highly covariant.}
 \label{tab:msp}
\end{deluxetable}

\subsubsection{PSR J1821+0155}

PSR J1821+0155 is unusual in that it has a 33 ms period, a spin down rate of 2.9$\times10^{-20}$ s s$^{-1}$, a weak magnetic field of $1.0\times 10^{9}$ G (see Table \ref{tab:slow_pulsars}), and no companion. \citet{lor04} interpreted such objects as being disrupted recycled pulsars (DRPs), later defined by \citet{bel10} as isolated pulsars having $B < 3\times 10^{10}$ G and $P > 20$ ms.

\citet{lor04} described a DRP as the first born neutron star in a binary system. The mass accretion from the companion decreases its period and magnetic field. The companion then undergoes a supernova explosion, disrupting the binary system, releasing both the neutron stars. The result is an isolated disrupted recycled pulsar and based on the properties of PSR J1821+0155, we classify it as a DRP. 

The ``survival probability" of a binary system to remain intact after the second supernova explosion is about 10\% \citep{lor04}. Therefore, DRPs are expected to be roughly 10 times as common as double neutron star (DNS) binaries in the Galaxy. However, observations indicate that there are comparable numbers of DRPs to DNS binaries. \citet{bel10} investigate this discrepancy using a sample of eight of the nine known DNS binaries and the 12 known DRPs\footnote{\citet{bel10} exclude 28 isolated pulsars with $B < 3\times 10^{10}$ G and $P < 20$ ms as they are believed to evolved from accretion from a low mass companion, such as a white dwarf, over a long period of time, resulting in the evaporation of the companion.}. They find that for an intermediate natal kick of $\sigma \sim170$ km s$^{-1}$, which results in a mean velocity of 99 and 154 km s$^{-1}$ for DNS and DRP populations respectively, there is a small ($\sim25$\%) observational bias toward detecting DNS binaries. The observational bias in favor of detecting DNS binaries is not enough to explain the discrepancy between the two populations, thus necessitating further study.

\begin{figure}
\begin{center}
\includegraphics[scale=1]{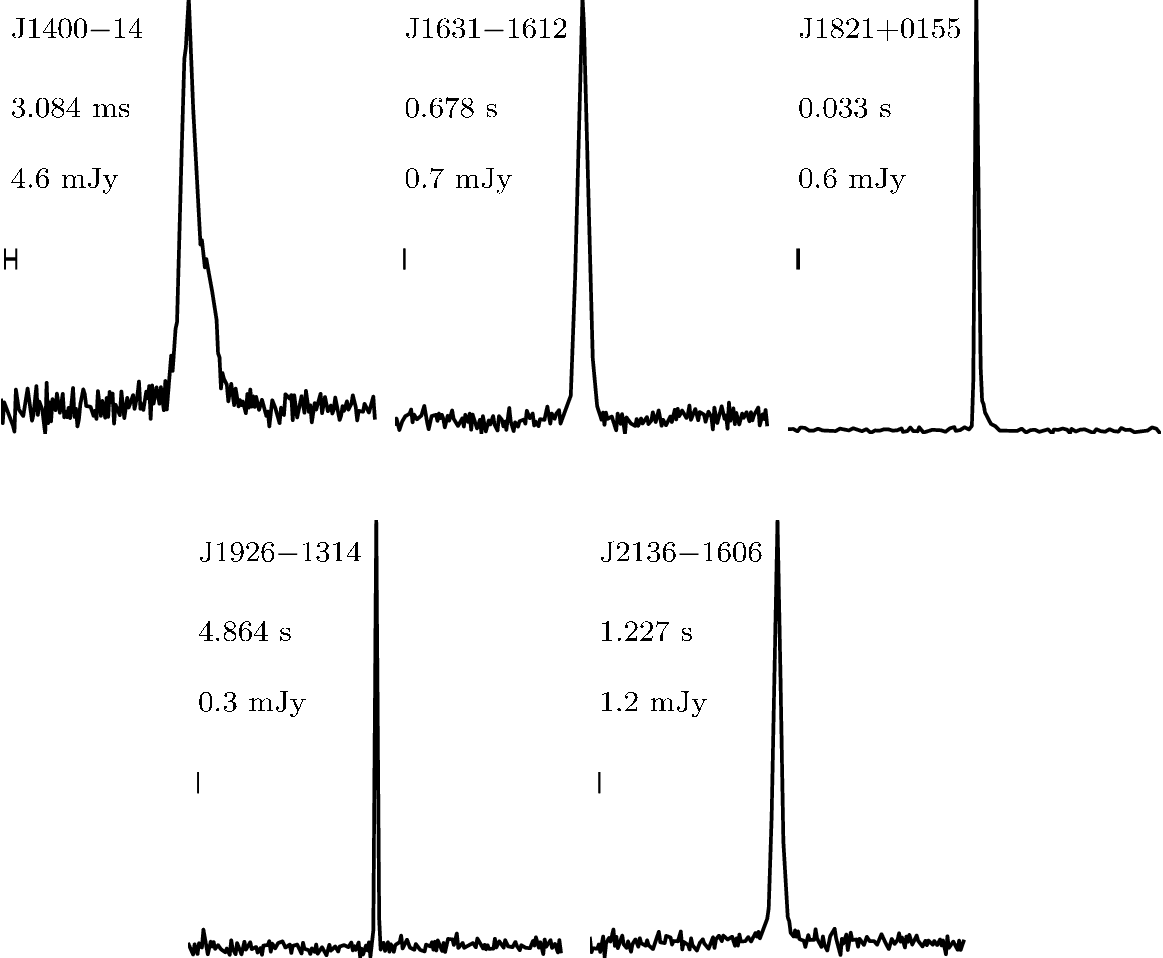}
\end{center}
\caption{The pulse profiles for the pulsars discovered by the students in the PSC drift-scan survey so far. All mean flux densities are from calibrated observations at 820 MHz, except for PSR J1400$-$1438, where the mean flux density is given for 350 MHz. Assuming a spectral index of $-1.7$, the equivalent flux density for PSR J1400$-$1438 at 820 MHz is 1.1 mJy. The integration times for PSRs J1631$-$1612, J1821+0155, J1926$-$1314, and J2136$-$1606 are 3.90, 5.89, 5.01 and 2.84 hours, respectively. The integration time for PSR J1400$-$1438 is 15 minutes due to the lack of observations at this frequency. The error bars on the left-hand side of each profile show the total expected DM smearing.}
\label{fig:slow_profs}
\end{figure}

We are continuing our timing observations of PSR J1821+0155. Because its profile is bright and narrow (see Figure \ref{fig:slow_profs}), we are able to get TOAs with about 2 $\mu$s errors for 5--10 minute observations at 820 MHz. Continued observations may allow us to measure a proper motion and estimate a spatial velocity of PSR J1821+0155. Because the difference in velocity distributions between DNS and DRP populations lead to different detection biases and the populations of each are small, a spatial velocity measurement of PSR J1821+0155 can provide a significant datapoint in further exploring the discrepancy between the predicted and observed population sizes.

\subsubsection{PSR J1926$-$1314}

The pulsar PSR J1926$-$1314 is notable due to its long period ($\sim4.9$ s, see Table \ref{tab:slow_pulsars}) and significant nulling fraction. Nulls are times when the radio emission ceases or is greatly reduced. To characterize the nulling fraction in PSR J1926$-$1314, we analyzed single pulses at both 350 MHz and 820 MHz, following a methodology similar to that of \citet{rit76}. After initially excising any radio frequency interference, we defined on-pulse and off-pulse windows, where on-pulse and off-pulse refer to regions of pulse phase based on integrated intensity. The on-pulse and off-pulse windows have the same number of bins, where the on-pulse window encompasses the entire pulse profile with as little baseline as possible. We then fit a second-order polynomial to the remaining bins and subtracted this from both the on-pulse and off-pulse windows to flatten the baseline. To account for scintillation, for each 200 pulse segment, we normalized the intensity by the average intensity of those 200 pulses ($I/[I]$). Any single pulses with an off-pulse intensity greater than four times the RMS of the off-pulse windows was removed.

\begin{figure}
\begin{center}
\includegraphics[scale=1]{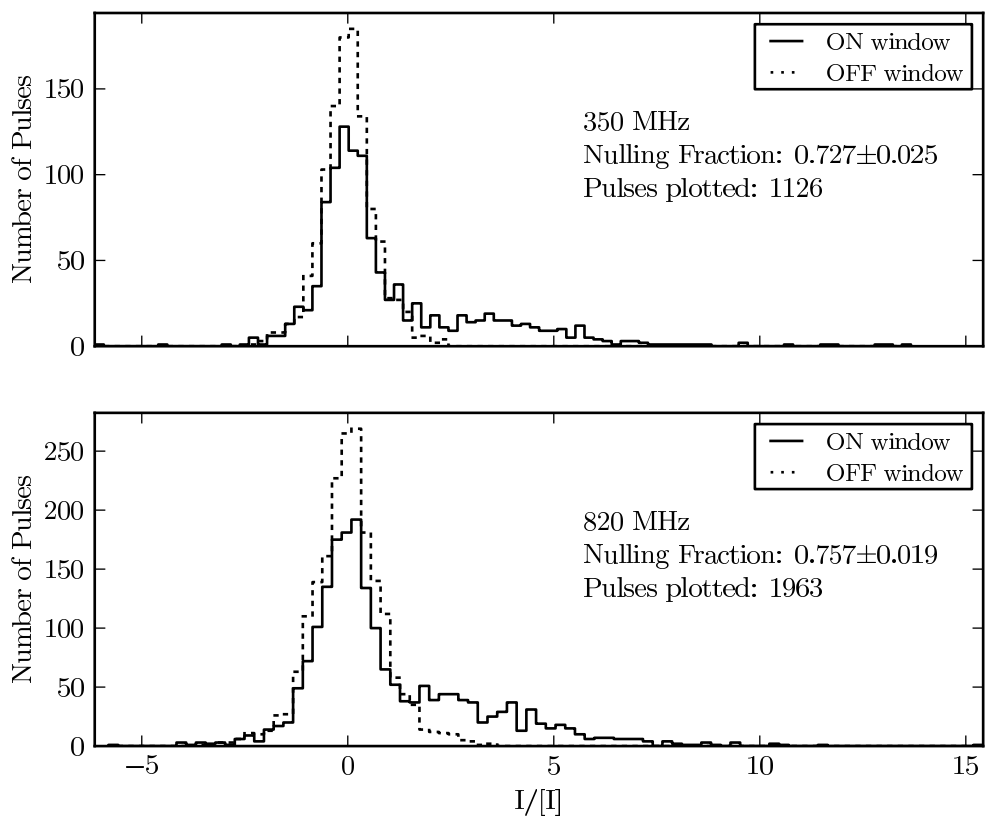}
\end{center}
\caption{A histogram of the single pulse intensity of the on-pulse and off-pulse windows at 350 MHz (left panel) and 820 MHz (right panel) for PSR J1926$-$1314. The x-axis is the normalized intensity, which accounts for scintillation effects, and is calculated by taking each pulse in a 200 pulse segment and subtracting the average intensity for those 200 pulses. The solid lines represent the on-pulse windows and the dashed lines represent the off-pulse windows. The on-pulse windows show a significant number of pulses with zero or negligible intensity indicating a large nulling fraction.}
\label{fig:nullfig}
\end{figure}

Figure \ref{fig:nullfig} shows a histogram of the single-pulse intensity for both the on-pulse and off-pulse windows at 350 MHz (left panel) and 820 MHz (right panel). To calculate the nulling fraction, we follow the methodology in \citet{wan07}, where we subtract the on-pulse distribution from the off-pulse distribution, which is multiplied by a trial nulling fraction starting with one and decreasing incrementally until the sum of the difference of the bins with intensities less than zero is equal to zero. The error in the nulling fraction is the square root of the number of null pulses divided by the total number of pulses. The observations at 820 MHz show a slightly higher nulling fraction than those at 350 MHz.

\citet{big92} analyzed 72 radio pulsars and found that pulsar period is directly proportional to nulling fraction, consistent with earlier work by \citet{rit76}, and suggested that older pulsars are harder to detect as they spend more time in their null state. \citet{wan07} found that nulling fraction correlates strongly with large characteristic age, even more than pulsar period, and \citet{cor08} showed that nulling fraction decreases with period derivative.

Because statistics on nulling pulsars are poor and the number of pulsars discovered in the GBT surveys is small, it is difficult to make any firm statements about the percentage of nullers in our survey compared to others. Of the 31 pulsars discovered in the GBT 350-MHz Drift Scan Survey and the PSC survey combined, three pulsars show a significant nulling fraction. By comparison, the Parkes Multibeam Survey \citep{man01} recognized significant nulling in 23 pulsars out of the 750 new discoveries of normal pulsars. This could be due to the PSC sky area coverage being away from the Galactic plane where pulsars are more likely to be older and show nulling behavior. In addition, the GBT 350-MHz Drift Scan Survey and the PSC survey are much more sensitive than previous surveys, allowing us to recognize nulling in more pulsars.

\section{Discussion}
\label{discussion}

One of the challenges of the PSC was the ability to manage the large number of diagnostic plots and to develop an interface for the students to quantitatively score the diagnostic plots. A further challenge was to collect all the scores, develop a metric for a given diagnostic plot, pseudo-pointing, student, or school, and to make these metrics easily accessible to the astronomers.  The PSC database was developed for this purpose. This database, which is freely available, is a new scheme for analyzing the diagnostic plots that is easily scalable and can be applied to other pulsar searches, benefiting both astronomers and students.

An important measure of the scientific success of the PSC is whether the students are able to detect pulsars as effectively as astronomers. The students have found 32 of 39 pulsars predicted to be detectable with S/N $\geq$~9 in the data they have examined so far and have re-detected an additional five known pulsars that were either below the S/N threshold or did not have any recorded flux measurements. Judging by detection rate, the students are on equal terms to
the astronomers in finding new pulsars. We expect there to be roughly 16 known pulsars and seven new pulsars in the remainder of the PSC data.

Of the new pulsars that the students have discovered, only one (PSR J1400$-$1438) is an MSP. PSR J1400$-$1438 will be considered for inclusion in a PTA because of its 1 $\mu$s TOA precision at 350 MHz, narrow pulse width, bright flux, and location in a sparsely MSP-populated region of the sky; it will not be added, however, unless it can be consistently detected/timed at multiple frequencies.  With an orbital period of 9.5 days and a projected semi-major axis of 8.4 light seconds, the companion is likely a white dwarf star. Based on the mass of the companion and the distance, we expect it to be a 19th magnitude star. This part of the sky is not covered by the Sloan Digital Sky Survey Data Release 9, but it is covered in the Original Digitized Sky Survey. No companion is visible in this survey. We are planning optical observations further try to identify the companion. PSR J1821+0155 is most likely a DRP and therefore it will likely have received a significant natal kick from its companion. We will be timing this pulsar to detect proper motion in the timing residuals. This pulsar will likely be considered in future studies of the detection rates of DRP and DNS systems. PSR J1926$-$1314 exhibits significant nulling and is one of three pulsars of the 31 pulsars discovered in the GBT 350-MHz Drift Scan Survey and the PSC surveys combined that show a significant nulling fraction, likely because these surveys are much more sensitive to an older pulsar population than previous surveys.

The PSC is continuing to operate with the fifth summer workshop having occurred in July 2012. We now have 76 teachers and 705 students from 18 states participating in the PSC. The students are currently examining the remaining PSC data. In addition, we have collected an additional 50 TB of data at 350 MHz and 820 MHz using GUPPI while the GBT is immobilized due to maintenance or inclement weather. These data will be processed in a similar fashion to the original PSC data, accounting for changes in elevation and observing frequency. These additional data will ensure that the PSC can continue through the growth of additional students and teachers over the coming years. Future plans include moving towards a distributed project in which students from many more high schools across the country can participate.

\acknowledgements

The National Radio Astronomy Observatory is a facility of the National Science Foundation operated under cooperative agreement by Associated Universities, Inc. We would like to acknowledge the NSF ITEST program for making the PSC possible. M.A.M. and D.R.L. are supported by a WVEPSCOR Research Challenge Grant and the Research Corporations.  We are grateful for the collaboration of all of the participants in the GBT 350-MHz Drift Scan Survey. We would like to thank P. Demorest for useful discussions. We would like to acknowledge the hard work and dedication of all the PSC teachers and their high schools for providing support to both the teachers and students in their participation in the PSC. 

\clearpage

\bibliographystyle{apj}

\end{document}